# Formation of Collapsed Tetragonal Phase in EuCo$_2$As$_2$ under High Pressure

**Matthew Bishop, Walter Uhoya, Georgiy Tsoi and Yogesh K. Vohra**

Department of Physics, University of Alabama at Birmingham (UAB), Birmingham, AL 35294, USA

**Athena S. Sefat and Brian C. Sales**

Oak Ridge National Laboratory (ORNL), Oak Ridge, TN 37831, USA

The structural properties of EuCo$_2$As$_2$ have been studied up to 35 GPa, through the use of x-ray diffraction in a diamond anvil cell at a synchrotron source. At ambient conditions, EuCo$_2$As$_2$ (*I*4/*mmm*) has a tetragonal lattice structure with a bulk modulus of 48 ± 4 GPa. With the application of pressure, the *a*-axis exhibits negative compressibility with a concurrent sharp decrease in *c*-axis length. The anomalous compressibility of the *a*-axis continues until 4.7 GPa, at which point the structure undergoes a second-order phase transition to a collapsed tetragonal (CT) state with a bulk modulus of 111 ± 2 GPa. We found a strong correlation between the ambient pressure volume of 122 parents of superconductors and the corresponding tetragonal to collapsed tetragonal phase transition pressures.

Introduction:

In recent years, there has been a continuous interest in using the pressure as a variable in the study and discovery of superconducting properties, magnetic and structural phase transitions in materials. Among such materials studied are the pnictides $A$Fe$_2$As$_2$ ($A$=divalent alkaline earth metal or divalent rare-earth metal). These form in the tetragonal ThCr$_2$Si$_2$ ('122')-type structure and can exhibit superconductivity under chemical-doping (isoelectronic- or electron-) and application of external pressure [1-5]. The Fe atom can be replaced by Co and the As atom can be replaced by the isoelectronic element P forming ternary phosphides in the same 122-type structure [6]. Studies on 122 phosphide systems have been carried out and isostructural transitions from tetragonal to collapsed tetragonal accompanied by negative compressibility axial properties have been reported in ternary phosphides under pressure [7, 8] and in other compounds [9]. It has been observed that the transition metal element essentially determines the nature of this phase transition in ternary phosphides. For instance, compounds such as EuFe$_2$P$_2$ exhibit a second order phase transition under pressure, however, in contrast to EuFe$_2$P$_2$ the pressure - induced structural phase transition is first-order in compounds such as EuCo$_2$P$_2$ in which Fe atom is replaced by Co atom [7, 8]. These experimental findings lead one to wonder whether the type of transition metal (T=Fe, Co) determines the first-order or second-order nature of tetragonal to collapsed-tetragonal phase transition in all 122-type compounds. Increase of interest in studying the compression behavior of the $A$Fe$_2$As$_2$ family have been generated by the recent discovery of superconductivity and also the pressure-induced isostructural (tetragonal to collapsed tetragonal) phase transition accompanied by extremely anisotropic and negative compressibility axial phenomenon in which the $a$-axis length increases while $c$-axis length decreases under pressure in the ternary iron pnictides superconductors; EuFe$_2$A$_2$ [10], BaFe$_2$As$_2$ and CaFe$_2$As$_2$ [11]. The negative compressibility phenomenon is suggested to be a common phenomenon in other iron based superconductors of type AT$_2$As$_2$ (A= Ba, Eu, Sr, Ca and T=Fe, Co). In order to investigate the structural properties of EuCo$_2$As$_2$, we have synthesized and performed measurement of the lattice parameters as a function of pressure at room temperature using synchrotron x-ray diffraction technique. We have been motivated by the occurrence of CT phase and exotic negative compressibility phenomenon in isostructural compounds of $ThCr_2Si_2$ type.

Experimental Details:

Single crystal samples of $EuCo_2As_2$ were grown from a CoAs flux, similar to that described in reference [2]. The crystals were ground into a polycrystalline sample and loaded into a 60-micron hole of a spring-steel gasket that was first pre-indented to a ~50 micron thickness and mounted in a diamond anvil cell for high pressure x-ray diffraction experiments. In this study, no pressure medium was employed and hence the structural transition reported in this study corresponds to a non-hydrostatic case.

The high pressure x-ray diffraction experiments were carried out at the beam-line 16-ID-B, HPCAT, Advanced Photon Source, Argonne National Laboratory. An angle dispersive technique with a MAR345 image-plate area detector was employed using a focused monochromatic beam with x-ray wavelength, $\lambda = 0.4325$ Å. Experimental geometric constraints and the sample to image plate detector distance were calibrated using a $CeO_2$ diffraction pattern and were held at the standard throughout the entirety of the experiment. Pressure was applied through the use of a diamond anvil cell and an internal copper pressure standard placed next to the sample was employed for the calibration of pressure .[12] The Birch Murnaghan equation [13] as shown by equation (1) was fitted to the available equation of state data on copper pressure standard [12].

$$P = 3B_0 f_E (1+2f_E)^{5/2} \left\{ 1 + \frac{3}{2}(B'-4)f_E \right\} \qquad (1)$$

Where $B_o$ is the bulk modulus, $B'$ is the first derivative of bulk modulus at ambient pressure, and $V_0$ is the ambient pressure volume. The fitted values for the copper pressure standard are $B_0 = 121.6$ GPa, $B' = 5.583$, and $V_0 = 11.802$ Å$^3$/atom. The parameter $f_E$ is related to volume compression and is described below.

$$f_E = \frac{\left[ \left(\frac{V_o}{V}\right)^{2/3} - 1 \right]}{2} \qquad (2)$$

Results and Discussions:

The diffraction images obtained were integrated using the program FIT2D [14] to yield intensity vs. diffraction angle (2 theta) plots. At ambient temperature and pressure, x-ray diffraction revealed a tetragonal structure with the lattice parameters $a = 3.9752 \pm 0.0031$ Å, $c = 11.1011 \pm 0.0032$ Å and an axial ratio $(c/a) = 2.7926 \pm 0.0032$ Å. These lattice parameters were confirmed independently through a collection of data using a PANalytical X'Pert Pro MPD at room temperature (using Cu K$\alpha$1 radiation), and a LeBail fit giving $a = 3.9671(1)$ Å and $c = 11.0632(5)$ Å. The ThCr$_2$Si$_2$-type structure has the following crystalline arrangement: Eu atoms at the 2a position (0,0,0), Co atoms at the 4d positions (0,1/2,1/4) and (1/2, 0, 1/4), and As atoms at the 4e positions (0,0,z) and (0,0,-z). The approximate structural parameter $z = 0.36$ has been obtained based on Rietveld refinement of x-ray diffraction data [15].

Figure 1 (a) shows the x-ray diffraction pattern at 0.33 GPa with EuCo$_2$As$_2$ sample in the *I*4/*mmm* tetragonal structure and the copper pressure marker in the face centered cubic (fcc) phase. The diffraction peaks are labeled with their respective (hkl) values. The tetragonal phase of EuCo$_2$As$_2$ is characterized by the (101), (110), (112), (200), (213) and (224) Bragg diffraction peaks. The fcc phase of the copper pressure marker is characterized by the (111), (200), (220), (311) and (222) Bragg diffraction peaks. The measured volume of copper pressure marker obtained from the x-ray diffraction was used to calculate the sample pressure from the equation of state given by equation (1). The obtained lattice parameters for the sample at a pressure of 0.33 GPa are $a = 3.9923 \pm 0.0031$ Å and $c = 10.7347 \pm 0.01$ Å. Figure 1(b) shows the x-ray diffraction pattern of the sample at 3.2 GPa with the lattice parameters $a = 4.0114 \pm 0.0031$ Å and $c = 10.0698 \pm 0.0194$ Å. When comparing the low-pressure spectrum figure 1(a) with figure 1(b), the sample peaks dependent only on the *a*-axis: (110) and (200) are all moving to lower diffraction angles while all non-*a*-dependent peaks move to higher diffraction angles and peaks dependent on both *a* and *c* remain approximately in the same position. Conventional behavior of the material under compression suggests that all peaks would move to higher angles in the x-ray diffraction spectrum; however, in comparing figure 1(a) with 1(b) all *a*-dependent peaks move to lower diffraction angles. This is a clear indication of negative compressibility in the *a*-axis of the tetragonal lattice structure in $EuCo_2As_2$ due to the increase in the *a*-axis concurrent with a sharp decrease in the *c*-axis with the application of pressure. Similar observations were seen in $EuFe_2As_2$, [10] $BaFe_2As_2$ and $CaFe_2As_2$ [11] during the transition to their respective collapsed tetragonal phases under compression. Another important note is that the predominant peaks unlabeled in the diffraction spectrums in figure 1, are identified as CoAs orthorhombic impurity phase (*Cmcm*) flux, which might have remained on the surface of the crystals. These peaks are predominant in spectrum below 13 GPa and become progressively less evident at higher pressures. Figure 1(c) shows that the sample remained in the tetragonal phase up to the highest pressure of 35.8 GPa with lattice parameters $a = 3.777 \pm 0.0021$ Å and $c = 9.2367 \pm 0.0493$ Å . The compression behavior above 4.7 GPa is considered normal as all Bragg peaks move to higher diffraction angles.

Figure 2 shows the measured *a*-axis as a function of pressure, exhibiting anomalous compression phenomenon. The *a*-axis shows an initial increase in axial length with increasing pressure up to 3.2 GPa prior to exhibiting normal compression up to a pressure of 35.8 GPa. The sample was then decompressed, exhibiting normal behavior up to 14 GPa where a hysteresis occurs. The *a*-axis begins to increase at a more rapid rate, reaching a maximum at 5.1 GPa prior to returning to slightly below its initial value. The anomalous behavior of the *a*-axis is paired with a sharp but continuous decrease in *c*-axis until 4.7 GPa, where a normal decrease in lattice parameter length occurs up to 35.8 GPa where upon decompression, the *c*-axis returned to approximately its initial length.

Figure 3 shows all the axial ratio (*c*/*a*) data points that were obtained during this experiment and we have combined the data obtained during compression and decompression in to one data set for subsequent analysis. The axial ratio shows a sharp decrease with increasing pressure up to 4.7 GPa and then a gradual decrease with further increase in pressure up to the highest pressure of 35.8 GPa. The variation of *c*/*a* with pressure can be divided in to two linear regions. The fits for the two linear regions are shown in figure 3 and are described by the following equations:

$$c / a = 2.7102 - 0.53P, \quad 0 \leq P \leq 4.7 \text{ GPa}, \quad (3)$$
$$c / a = 2.4762 - 0.001P, \quad 0 \leq P \leq 35.8 \text{ GPa} \quad (4)$$

The intersection of the two linear regions as described by equations (3) and (4) defines the phase transition from the ambient pressure T phase to the CT phase. $EuCo_2As_2$ is found to undergo this phase transition at 4.7 GPa at ambient temperature. The nature of this phase transition is clearly continuous as can be seen in pressure dependence of the structural parameters (figures 2, 3 and 4). Similar continuous T to CT phase transition has been observed in the compression behavior of the related $ThCr_2Si_2$ structure-type compounds of $EuFe_2As_2$, [10] $EuFe_2P_2$ and $LaFe_2P_2$ [8]. This is in sharp contrast to observations made in ternary phosphides in which replacing Fe atom with Co atom reduces the nature of the observed structural phase transition from second order (continuous) to first order (discontinuous) type [7,8]. The bonding characteristics of As and P atoms along with the pressure induced changes in electronic structure determine the formation of collapsed tetragonal structure in these materials. . An additional consideration for present studies is that the high-pressure Mössbauer investigations on $EuFe_2P_2$ have revealed a continuous structural phase transition which is accompanied with a continuous valence transition from a magnetic $Eu^{2+}$ state to a nonmagnetic $Eu^{3+}$ state in the pressure range between 3-9 GPa [16]. We expect this valence fluctuation to have an influence on magnetic phase transitions at low temperatures and have an overall effect on structural phase transitions under high pressure at room temperature due to enhanced compressibility that accompanies valence fluctuations.

Figure 4 shows the measured pressure-volume curve of $EuCo_2As_2$ up to 35.8 GPa at ambient temperature. The pressure -volume data were fitted by the Birch-Murnaghan equation of state described by equation (1). The fits from the axial ratio showed a phase transition from the T to the CT phase at 4.7 GPa. The transition pressure obtained from the axial ratio was used to separate the tetragonal phase from the collapsed tetragonal phase when determining the equation of state for $EuCo_2As_2$. The fitted zero-pressure volumes ($V_0$) for the tetragonal and collapsed tetragonal phases are 172.4 ± 0.7 $Å^3$ and 165.9 ± 0.3 $Å^3$ respectively. The measured change in volume with the application of pressure shows a significant decrease in overall unit cell volume indicating the collapsed tetragonal phase has a 3.8% higher density than the tetragonal phase at zero pressure. The Birch-Murnaghan fit revealed a bulk modulus for the tetragonal phase to be $B_o = 48 ± 4$ GPa and 111 ± 2 GPa for the CT phase. This comparison shows that the T phase is 2.3 times more compressible than the CT phase at zero pressure. Due to the limited number of data points in the tetragonal phase, the first derivative of the bulk modulus could not be estimated and was fixed at 4. The calculated first derivative of the bulk modulus for the collapse tetragonal phase was found to be $B` = 3.06$. These fit parameters are summarized in Table 1.

Figure 5 shows the correlation between the observed zero pressure volume ($V_0$) for $AT_2As_2$ 122 materials ($A$=Ba ,Fe, Ca and Eu ;T=Fe, Co) and their corresponding phase transition pressures ($P_T$) from T to CT phases at 300 K. The data for $EuFe_2As_2$ were obtained from ref [10] and the data for $CaFe_2As_2$ were taken from ref [11]. The variation shows a nearly linear increase in $P_T$ with increasing $V_0$. The solid curve is the linear fit to data and is described by the equation below:

$$P_T = 0.388 V_0 (+/-0.034) \text{ GPa}/\text{Å}^3 - 63.259 (+/- 6.2) \text{ GPa}. \qquad (5)$$

Following this work, we suggest that transition from the T- to CT-phase, under compression with a concurrent negative axial compressibility, is a common effect in all the parents of iron-based arsenic 122 superconductors and perhaps other related compounds of the 122-type, The T- to CT-phase of these compounds at 300 K can be estimated using equation (5) if their corresponding zero pressure volume at ambient temperature is known. The pressure induced isostructural phase transition to a collapsed tetragonal at ambient temperature is not unique among arsenic compounds that have $ThCr_2Si_2$ type structure. The effect is found among the phosphides that are isostructural to $AT_2As_2$ compounds such as $EuCo_2P_2$ and $SrNi_2P_2$ [7], $EuFe_2P_2$ [16]. These phosphides compounds have been widely studied and the rapid decrease observed in their c/a ratios in the collapsed tetragonal phase has been attributed to bonding transitions associated with the formation of P-P single bond between ions in adjacent planes along the c-axis [17]. In connection to arsenic compounds, recent theoretical calculations for $CaFe_2As_2$ suggest that there is similar transition in the bonding character of As ions and the enhancement of the As-As bonds across the $Fe_2As_2$ layers under pressure [18]. This work and previous studies on the related 122 compounds suggest that the collapsed tetragonal phase, with the space group.*I4/mmm*, is the stable high-pressure phase of the $AT_2As_2$ type compounds, and there have been no any evidence of post collapsed phase transitions at least up to maximum pressures of these studies (70 GPa in $EuFe_2As_2$ [10], 56 GPa in $BaFe_2As_2$, 51 GPa in $CaFe_2As_2$ [11] and 35 GPa $EuCo_2As_2$).

Conclusions:

In summary, we have studied the pressure effects on the crystal structure of the layered $ThCr_2Si_2$-type $EuCo_2As_2$ up to 35.8 GPa at ambient temperature using synchrotron x-ray diffraction. The x-ray diffraction patterns collected reveal a highly anisotropic and anomalous compressibility effects in which *a*-axis increases with increasing pressures up to a maximum then decreases normally while the *c*-axis decreases continuously with increasing pressure. Analysis of the x-ray diffraction data indicates a tetragonal phase to a collapsed tetragonal phase transition in $EuCo_2As_2$ at 4.7 GPa. The equations of state for the T- and CT-phases show distinct Bulk Moduli. At ambient pressure, an extrapolated CT phase has a density that is 3.8 % higher as compared to the T-phase under similar conditions. We suggest that the transition from T- to CT-phase under compression, along with a concurrent negative axial compressibility, is a common effect in every compound of the type $AT_2As_2$ ($A$ = Ba, Ca, Sr, Eu and T = Fe, Co). A linear correlation between the ambient pressure volume for such 122 materials and their corresponding phase transition pressures from T- to CT-phase at 300K is obtained. Additional systematic theoretical and experimental studies of structural phase transitions in $AT_2As_2$ compounds under hydrostatic and non-hydrostatic pressure conditions are required to clearly establish a correlation between transition pressure, initial volume, and electronic structure of these materials

ACKNOWLEDGMENT

Matthew Bishop acknowledges support from the National Science Foundation (NSF) Research Experiences for Undergraduates (REU)-site under Grant No. NSF-DMR-06446842. Walter Uhoya acknowledges support from the Carnegie/Department of Energy (DOE) Alliance Center (CDAC) under Grant No. DE-FC52-08NA28554 Research at Oak Ridge National Laboratory is sponsored by the Materials Sciences and Engineering Division, Office of Basic Energy Sciences, US Department of Energy.

**Figure Captions:**

Fig. 1:

The integrated x-ray diffraction profiles for $EuCo_2As_2$ and copper pressure standard at various pressures and at 300 K recorded with x-ray wavelength $\lambda = 0.4325$ Å. The Miller indices are indicated for $EuCo_2As_2$ tetragonal phase and face centered cube phase for copper. (a) Sample is in the ambient pressure T-phase at a pressure of 0.3 GPa, (b) sample is in T- phase at the height of negative compressibility discussed in the text at a pressure of 3.2 GPa, and (c) sample is in CT- phase at the highest pressure of 35.8 GPa. All spectrums contain extra diffraction peaks from CoAs flux, contamination on surface of ground crystals, with space group number #62.

Fig. 2:

The *a*-axis length dependence on pressure, with negative compressibility effects reaching a maximum at about 3.2 GPa and exhibiting a normal compression behavior beyond and up to 35.8 GPa. The decompression of the sample from high pressure shows a hysteresis in the *a*-axis beginning at approximately 14 GPa relaxing at a maximum of 5.0 GPa prior to returning to approximately its original axial length. The error bars are smaller than the size of the symbol used in plotting.

Fig. 3:

The measured axial ratio (*c/a*) for the tetragonal phase of $EuCo_2As_2$ as a function of pressure to 35.8 GPa at 300 K. The plot contains data obtained during both compression and decompression measurements. The linear fits for the two phases, i.e., the Tetragonal (T) phase and the Collapsed Tetragonal (CT) phase are described in the text. The intersection between the two linear fit lines defines the C-phase to CT-phase transition pressure and occurs at 4.7 GPa.

Fig. 4:

The pressure dependence of the unit cell volume compression (V) for the tetragonal phase of $EuCo_2As_2$ up to a pressure of 35.8 GPa and at 300 K. The solid curves are the Birch Murnaghan equation of state fit to the two phases, i.e., the Tetragonal (T) phase and the Collapsed Tetragonal (CT) phase. The fit parameters are summarized in Table 1. The transition between the two phases occurs at 4.7 GPa.

Fig. 5:

The correlation between the ambient pressure volume for $A$T$_2$As$_2$ 122 materials and their corresponding phase transition pressures from tetragonal (T) to Collapsed Tetragonal (CT) phases at 300 K. The solid curve is the linear fit to data and is described in the text. The data for EuFe$_2$As$_2$ were obtained from ref [10] and the data for CaFe$_2$As$_2$ were taken from ref [11].

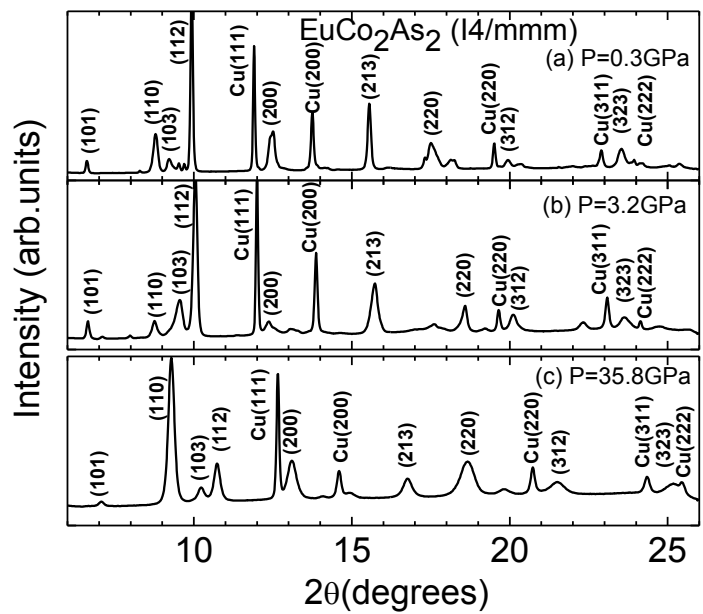

Figure 1

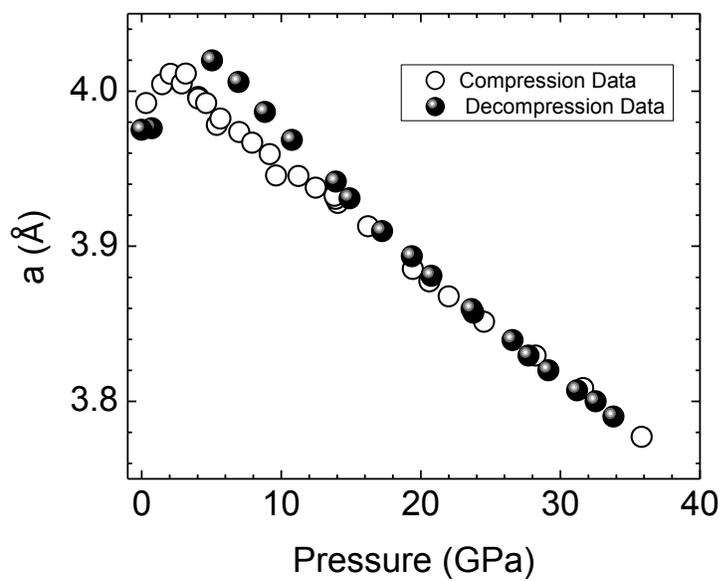

Figure 2

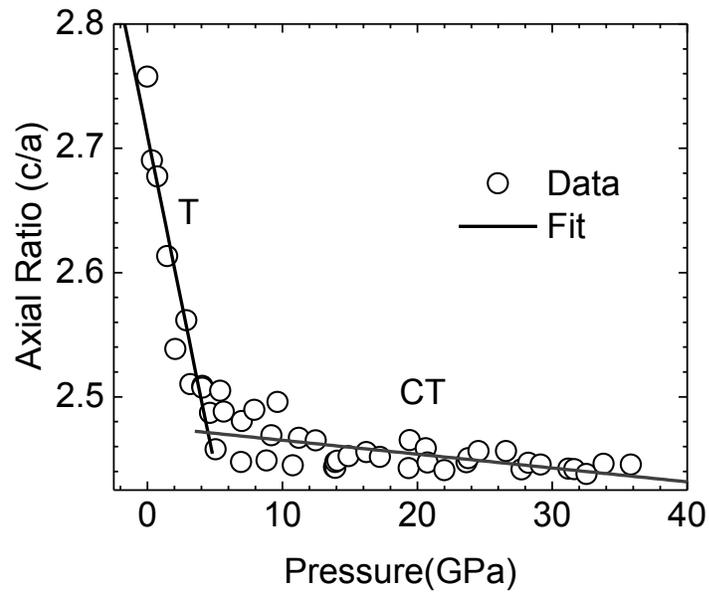

Figure 3

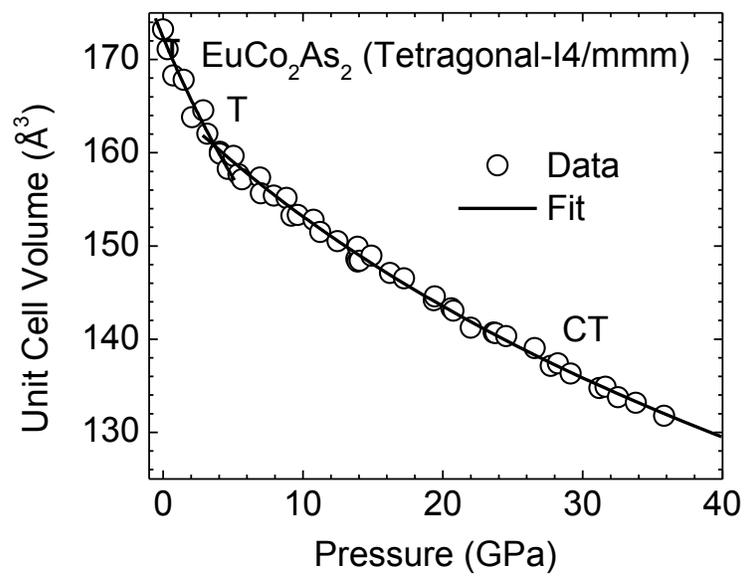

Figure 4

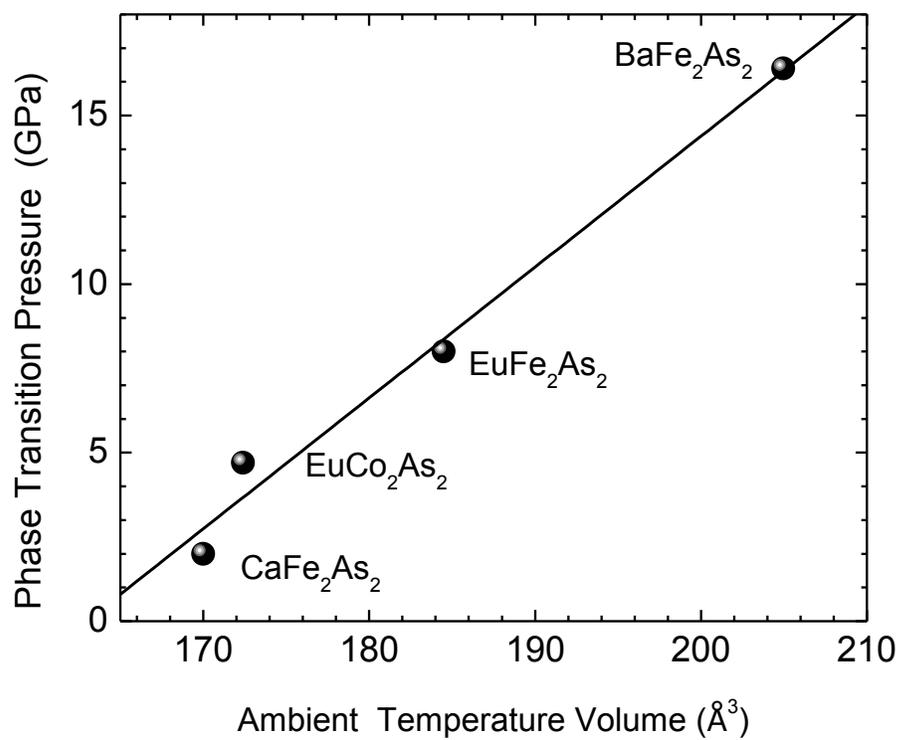

Figure 5

Table 1: Equation of state parameters for the $EuCo_2As_2$ in the Tetragonal (T) phase and Collapsed Tetragonal (CT) phase at ambient temperatures

| Phase | Unit Cell Volume ($V_0$) at Ambient Conditions | Bulk Modulus ($B_0$) | Pressure Derivative of Bulk Modulus ($B´$) |
|---|---|---|---|
| Tetragonal (T) 0 < P < 4.7 GPa | 172.4 ± 0.7 Å$^3$ | 48 ± 4 GPa | 4(assumed) |
| Collapsed Tetragonal (CT) 5< P < 35.8 GPa | 165.9 ± 0.3 Å$^3$ | 111 ± 2 GPa | 3.06 |